\documentclass[aps,showpacs,amsmath,amssymb,reprint]{revtex4-1} 
\usepackage{mathrsfs}
\usepackage{array}
\usepackage{amsmath}
\usepackage{graphicx}
\usepackage{amstext}
\usepackage{amsfonts}
\usepackage{bm}
\usepackage{hyperref}

\begin{document}
\title{Intrinsic probability distributions for physical systems}
\author{Tzu-Chao Hung}
\email{tchung618@gmail.com} 
\affiliation{Department of Physics, National Cheng Kung University, Tainan 70101, Taiwan}

\begin{abstract}
For a given metric $g_{\mu\nu}$, which is identified as Fisher
information metric, we generate new constraints for the probability
distributions for physical systems. We postulate the existence of
intrinsic probability distributions for physical systems, and
calculate the probability distribution by optimizing the Fisher
information metric under specified constraints. Accordingly, we get
differential equations for the probability distributions.
\end{abstract}
\maketitle

\section{The Fisher Information Metric}

Fisher information proposed by Fisher \cite{Fisher} is a way to
estimate hidden parameters in a set of random variables. Since we want to retrieve information of certain parameters $\xi_i$ in a statistical set for random variables or added noise $x_i$.  A measurement $y_i$ of the parameters have the relation with $\xi_i$ and $x_i$ write
\begin{equation*}
y_i=\xi_i+x_i.
\end{equation*}
Accordingly, we assume the likelihood $\rho(y_i|\xi_i)$ will have the relation that
\begin{equation*}
\rho(y_i|\xi_i)=\rho(y_i-\xi_i)=\rho(x_i),
\end{equation*}
which can be considered that $y_i$ is measurements of the positions of a particle, and $\xi_i$ is the actual positions of the particle.  Therefore, the Fisher information matrix is defined by~\cite{Nagaoka}
\begin{eqnarray*}
I_{kl}&=&\int d\mu(y_i)\,\rho(y_i|\xi_i) \frac{\partial\ln \rho(y_i|\xi_i)}
{\partial\xi^k}\frac{\partial\ln \rho(y_i|\xi_i)}{\partial\xi^l}\nonumber\\
&=&\int d\mu(x_i)\, \frac{1}{\rho(x_i)}\frac{\partial\rho(x_i)}
{\partial x^k}\frac{\partial\rho(x_i)}{\partial x^l}\ ,
\end{eqnarray*}
where $\int d\mu(x_i)$ integrates the whole space.  Let $\rho(x_i)=\Psi^2(x_i)$, where $\Psi$ is real.  Then the
Fisher information matrix can be rewritten as
\begin{equation*}
I_{kl}=4\int d\mu(x_i)\frac{\partial\Psi(x_i)}{\partial x^k}
\frac{\partial\Psi(x_i)}{\partial x^l}\ ,
\end{equation*}
which is symmetric, $I_{kl}=I_{lk}$.

Fisher information matrix also provides a natural distinguishability
metric for probability distributions which can be expressed by
\cite{Caves,Wootters}
\begin{equation*}
ds_{\mathrm{PD}}^2\equiv\sum_j\frac{d\rho_j^2}{\rho_j}=4\sum_jd\Psi_j^2.
\end{equation*}
Wootters called $s_{\mathrm{PD}}$ the statistical distance, which is
defined by ``maximum number of mutually distinguishable intermediate
probabilities.''  Hence, the Fisher information matrix can be
generalized to Fisher information metric, which is a Riemannian
metric on a smooth manifold.  Due to the Fisher information metric determines the maximum number of distinguishable probabilities.  So we postulate that to extreme the Fisher information can give the probability distributions in a physical system.

\section{Probability Distributions in Minkowski Space}

We can use a metric $g_{\mu\nu}$ to describe the geometric
properties of a space-time.  If there exist probability
distributions intrinsically, then, we expect that the motion of tiny particles will be described by the distributions.  In the other words, and the main idea in this article, we postulate that a certain geometry in a physical system exist a certain configurations of probability distributions or wave functions.  By identifying the metrics for physical space with the Fisher information metric, which determines the maximum number of distinguishable probabilities, and choosing the constraints, such as normalization and localization of probabilities, via Lagrange multipliers.  Therefore, according to the method which is proposed by Frieden called extreme physical information (EPI), then we can find the probability distributions for physical system~\cite{Frieden2004}.  And we suggest to call it intrinsic probability distributions for physical systems.

\subsection{Minkowski space in spherical coordinates}

The Minkowski metric in spherical coordinates is $\eta_{\mu\nu}=(c^2,1,r^2,r^2\sin^2\theta)$~\cite{Minkowski}.  The line element squared reads
\begin{equation*}
ds^2=-d\tau^2+dr^2+r^2d\theta^2+r^2\sin^2\theta d\phi^2,
\end{equation*}
where $d\tau=cdt$, and $c$ is the speed of light.  We defined the Fisher information metric as follows~\cite{Calmet} :
\begin{equation*}
I_{\mu\nu}=\eta_{\mu\nu}.
\end{equation*}
The function $\Psi$ could be considered that it is composed by four estimate parameters or , briefly, variables $\tau$, $r$, $\theta$, $\phi$, expressed by $\Psi=\Psi(\tau , r , \theta, \phi)$.  And with the normalization condition $\int d\mu(x_i)\, \Psi^2=1$.  We rewrite the diagonal terms of Fisher information metric
\begin{eqnarray*}
\int d\mu(x_i)\,\left(\frac{\partial \Psi}{\partial \tau}\right)^2 &=& -\frac{1}{4}, \\
\int d\mu(x_i)\,\left(\frac{\partial \Psi}{\partial r}\right)^2 &=& \frac{1}{4}, \\
\int d\mu(x_i)\,\left(\frac{\partial \Psi}{\partial \theta}\right)^2 &=& \frac{1}{4}r^2, \\
\int d\mu(x_i)\,\left(\frac{\partial \Psi}{\partial \phi}\right)^2 &=& \frac{1}{4}r^2\sin^2\theta. 
\end{eqnarray*}
The volume integral $\int d\mu(x_i) $ is given by $\int\sqrt{-\eta}\,d\tau dr d\theta d\phi$.  We assume that the right-hand side of the diagonal terms of Fisher information metric are representing the expectation values, which expressed by $\langle {\rm A} \rangle = \int d\mu(x_i)\,{\rm A}\Psi^2$.  Therefore, we have
\begin{eqnarray}
\int d\mu(x_i)\,\left[ -\kappa_0^2\Psi^2 - \left(\frac{\partial \Psi}{\partial \tau}\right)^2\right] &=& 0, \label{FIMtau} \\
\int d\mu(x_i)\,\left[ \kappa_1^2\Psi^2 - \left(\frac{\partial \Psi}{\partial r}\right)^2\right] &=& 0, \\
\int d\mu(x_i)\,\left[ \kappa_2^2 r^2 \Psi^2 - \left(\frac{\partial \Psi}{\partial \theta}\right)^2\right] &=& 0, \\
\int d\mu(x_i)\,\left[ \kappa_3^2 r^2\sin^2\theta \Psi^2 - \left(\frac{\partial \Psi}{\partial \phi}\right)^2\right] &=& 0 \label{FIMphi}, 
\end{eqnarray}
where the factor $1/4$ is absorbed into the Lagrange multipliers $\kappa_i$.  We note that the unit vectors in the spherical polar coordinates are not constant, they depend on the position vector.  From the constraints of Fisher information metric, Eq.~(\ref{FIMtau}-\ref{FIMphi}), we can determine the gradients in the spatial dimensions are:
\begin{eqnarray*}
\nabla_\tau\Psi=-i\dfrac{\partial\Psi}{\partial\tau}\hat{\tau}&,& \,\,
\nabla_r\Psi=\frac{\partial\Psi}{\partial r}\hat{r}, \\
\nabla_\theta\Psi=\frac{1}{r}\frac{\partial\Psi}{\partial\theta}\hat{\theta}&,& \,\,
\nabla_\phi\Psi=\frac{1}{r\sin\theta}\frac{\partial\Psi}{\partial\phi}\hat{\phi}. \nonumber
\end{eqnarray*}
Therefore we redefine the constraints of Fisher information metric, which are $\int d\mu(x_i)\,(\nabla_\tau\Psi)^2=\langle\kappa_0^2\rangle$, $\int d\mu(x_i)\,(\nabla_r\Psi)^2=\langle\kappa_1^2\rangle$, $\int d\mu(x_i)\,(\nabla_\theta\Psi)^2=\langle\kappa_2^2\rangle$ and $\int d\mu(x_i)\,(\nabla_\phi\Psi)^2=\langle\kappa_3^2\rangle$.  With a setted metric, we determine the rule of gradient and define the constraints which the probability distributions must satisfy.  In order to obtain the probability distributions by using the method called extreme physical information, we determine the objective function for the Fisher information metric for variations that vanish at the boundary, reads 
\begin{equation}
{\mathscr L}=\left(\nabla_\tau\Psi\right)^2+\left(\nabla_r\Psi\right)^2 +\left(\nabla_\theta\Psi\right)^2 +\left(\nabla_\phi\Psi\right)^2-\alpha^2\Psi^2\ , \label{Lalpha}
\end{equation}
where $\alpha^2=\sum_{i=0}^3\kappa_i^2$.
\begin{widetext}
The Euler-Lagrange equation reads
\begin{equation}
\dfrac{\partial^2\Psi}{\partial\tau^2}-\left[\frac{1}{r^2}\frac{\partial}{\partial r} \left(r^2\frac{\partial\Psi}{\partial r}\right) +\frac{1}{r^2\sin\theta}\frac{\partial}{\partial\theta}
\left(\sin\theta\frac{\partial\Psi}{\partial\theta}\right) +\frac{1}{r^2\sin^2\theta}\left(\frac{\partial^2\Psi}{\partial\phi^2}\right)\right]
=\alpha^2\Psi \ .\label{HE01}
\end{equation}
Considering the problem of finding solution of the form $\Psi(\tau, r,\theta,\phi) = T(\tau)
R(r)\Theta(\theta)\Phi(\phi)$.  By separation variables, we obtain four differential equations
\begin{eqnarray}
\dfrac{d^2T(\tau)}{d\tau^2}+\eta^2T(\tau)&=&0,\\
\dfrac{1}{r^2}\dfrac{d}{d r}\left(r^2\dfrac{dR(r)}{d r}\right)+\left(\alpha^2+\eta^2-\dfrac{\ell(\ell+1)}{r^2}\right)R(r)&=&0, \label{radialEQ}\\
\dfrac{1}{\sin\theta}\dfrac{d}{d \theta}\left(\sin\theta\dfrac{d\Theta(\theta)}{d \theta}\right)+\left[ \ell(\ell+1)-\dfrac{m^2}{\sin^2\theta}\right]\Theta(\theta)&=&0,\label{polarEQ}\\
\dfrac{d^2\Phi(\phi)}{d\phi^2}+m^2\Phi(\phi)&=&0.\label{azimuthalEQ}
\end{eqnarray}
The solution, $\Psi(\tau,r,\theta,\phi)$, is
\begin{eqnarray*}
\Psi(\tau,r,\theta,\phi) = Aj_\ell(\alpha r)\epsilon\sqrt{\dfrac{(2\ell+1)}{4\pi}\dfrac{(\ell-|m|)!}{((\ell+|m|)!)}} P_\ell^m(\cos\theta)e^{im\phi}e^{-i\eta\tau},
\end{eqnarray*}
where $\epsilon=(-1)^m$ for $m\geq 0$, and $\epsilon=1$ for $m\leq
0$, $\ell=0,1,2,...$, $|m| \leq \ell$, $\eta=0,\pm 1, \pm 2, \pm 3,...$, $A$ is a normalization constant, $j_\ell(x)$ is the spherical
Bessel function of order $\ell$, and $P_\ell^m$  the associated
Legendre functions. 

Now we consider the local property in radial estimation and choosing the reference point at the origin. For an optimal measurement, we consider that
\begin{equation*}
\dfrac{1}{\sigma_r^2}\int d\mu(x_i)\, r^2 \Psi^2 =1.
\end{equation*}
The objective function for the  Fisher information metric
\begin{equation}
{\mathscr L}=\left(\nabla_\tau\Psi\right)^2\left(\nabla_r\Psi\right)^2
+\left(\nabla_\theta\Psi\right)^2
+\left(\nabla_\phi\Psi\right)^2+\beta^2 r^2\Psi^2-\alpha^2\Psi^2\ , \label{Lkappa}
\end{equation}
where $\beta^2$ is a multiplier. The Euler-Lagrange equation reads
\begin{equation}
\dfrac{\partial^2\Psi}{\partial\tau^2}-\left[\frac{1}{r^2}\frac{\partial}{\partial r} \left(r^2\frac{\partial\Psi}{\partial r}\right) +\frac{1}{r^2\sin\theta}\frac{\partial}{\partial\theta}
\left(\sin\theta\frac{\partial\Psi}{\partial\theta}\right) +\frac{1}{r^2\sin^2\theta}\left(\frac{\partial^2\Psi}{\partial\phi^2}\right)\right]
+\beta^2 r^2\Psi=\alpha^2\Psi \ .\label{HE}
\end{equation}
Considering the problem of finding solution of the form $\Psi(\tau, r,\theta,\phi) = T(\tau)R(r)\Theta(\theta)\Phi(\phi)$.  By separation variables, we obtain four differential equation
\begin{eqnarray}
\dfrac{d^2T(\tau)}{d\tau^2}+\eta^2T(\tau)&=&0,\\
\dfrac{1}{r^2}\dfrac{d}{d r}\left(r^2\dfrac{dR(r)}{d r}\right)+\left(\alpha^2-\beta^2 r^2-\dfrac{\ell(\ell+1)}{r^2}\right)R(r)&=&0, \label{radialEQ'}\\
\dfrac{1}{\sin\theta}\dfrac{d}{d \theta}\left(\sin\theta\dfrac{d\Theta(\theta)}{d \theta}\right)+\left[ \ell(\ell+1)-\dfrac{m^2}{\sin^2\theta}\right]\Theta(\theta)&=&0,\label{polarEQ'}\\
\dfrac{d^2\Phi(\phi)}{d\phi^2}+m^2\Phi(\phi)&=&0.\label{azimuthalEQ'}
\end{eqnarray}
The solution, $\Psi(\tau,r,\theta,\phi)$, reads
\begin{eqnarray*}
\Psi(\tau,r,\theta,\phi)=B\, 2^{\frac{1}{4}(1-2\ell)}r^{-(1+\ell)}e^{-\frac{\beta r^2}{2}} L_{\frac{\alpha^2}{4\beta}-\frac{1}{4}+\frac{\ell}{2}}^{-(\frac{1}{2}-\ell)}(\beta r^2)
\times\epsilon\sqrt{\dfrac{(2\ell+1)}{4\pi}\dfrac{(\ell-|m|)!}{((\ell+|m|)!)}}
P_\ell^m(\cos\theta)e^{im\phi}e^{-i\eta\tau},
\end{eqnarray*}
where $\epsilon=(-1)^m$ for $m\geq 0$, and $\epsilon=1$ for $m\leq
0$, $\ell=0,1,2,...$, $|m| \leq \ell$, $\eta=0,\pm 1, \pm 2, \pm 3,...$, $B$ is a normalization constant
and
\begin{equation*}
L_{\frac{\alpha^2}{4\beta}-\frac{1}{4}+\frac{\ell}{2}}^{-(\frac{1}{2}-\ell)}(\beta r^2)
\end{equation*}
are the generalized Laguerre polynomials.
\end{widetext}

\subsection{Connection to quantum physics}
From above derivations, we find that the physical system is specified by configurations of probability distributions according to Fisher
information metric. Although we have the differential equation,
Eq.~(\ref{HE01}), to describe the probability distribution in the
spherical Minkowski space, we still want to understand the phenomena
when we put a test particle in the space. Therefore, we express the
function $\Psi(\tau,r,\theta,\phi)$ in its momentum space by Fourier
transform, as~\cite{Frieden2004}
\begin{eqnarray}
\Psi(\tau,r,\theta,\phi)&=&\dfrac{1}{(2\pi\hbar)^{3/2}}\int dE dp_rdp_\theta dp_\phi \Phi(E,p_r,p_\theta,p_\phi)\nonumber\\ 
& & \times e^{-i(-E\tau+p_r r+p_\theta \theta+p_\phi \phi)/\hbar}, \label{project01}
\end{eqnarray}
where we note that the dimension of Planck's constant $h$ is same as
angular momentum's, $\hbar = h/2\pi$. We substitute
Eq.~(\ref{project01}) into Eq.~(\ref{Lalpha}) and get the Lagrangian
as $\mathscr{L}=\left(-\frac{1}{\hbar^2}E^2+\frac{2\mu}{\hbar^2}E_{\rm
kr}+\frac{L_\theta^2}{\hbar^2}+\frac{L_\phi^2}{\hbar^2}-\alpha^2\right)\Psi^2,$
where $E$ is the total energy of the particle, $p_\theta$ is the canonical momentum to $\theta$, $L_\theta$
is the angular momentum for the variable $\theta$, $p_\phi$ is the
canonical momentum to $\phi$, and $L_\phi$ is the angular momentum
for the variable $\phi$.  The kinetic energy in radial orientation
$E_{\rm kr} $ equals to $p_r^2 /2\mu$, where $\mu $ is the reduced
mass of the test particle.  A particle with non-zero
rest mass, $\mu\neq 0$, must satisfy a conservation equation, which
is~\cite{MTW}
\begin{equation*}
\eta_{\lambda\nu}p^\lambda p^\nu + \mu^2c^2=\eta^{\lambda\nu}p_\lambda p_\nu + \mu^2c^2=0,
\end{equation*}
where $c$ is the speed of light.  Therefore, we determine that
$\alpha^2$ should equal to $-\mu^2c^2/\hbar^2$, a fixed condition to fulfil the required constraint equation, then we put a test particle in the physical system, we see that the behavior of the test particle obeys the Klein-Gordon equation.

\section{Probability Distribution in Schwarzschild Geometry}

The Schwarzschild metric~\cite{Schwarzschild1916} has the form
$g_{\mu\nu}=(-(1-r_s/r),\,(1-r_s/r)^{-1},\,r^2,\,r^2\sin^2\theta)$
for the four dimensional coordinates $(\tau,r,\theta,\phi)$ with
$r_s=2GM/c^2 $ being the Schwarzschild or gravitational radius,
where $G$ is the gravitational constant.  The invariant line element
squared is given by
\begin{equation*}
ds^2=-\left(1-\frac{r_s}{r}\right)d\tau^2+\left(1-\frac{r_s}{r}\right)^{-1}dr^2
+r^2d\theta^2+r^2\sin^2\theta d\phi^2,
\end{equation*}
where $d\tau=cdt$, and $c$ is the speed of light.  We rewrite the
diagonal terms of Fisher information metric
\begin{eqnarray*}
\int d\mu(x_i)\,\left(\frac{\partial \Psi}{\partial \tau}\right)^2&=&-\frac{1}{4} \left(1-\frac{r_s}{r}\right),\\
\int d\mu(x_i)\,\left(\frac{\partial \Psi}{\partial r}\right)^2&=&\frac{1}{4}\left(1-\dfrac{r_s}{r}\right)^{-1},\\
\int d\mu(x_i)\,\left(\frac{\partial \Psi}{\partial \theta}\right)^2&=&\frac{1}{4} r^2,\\
\int d\mu(x_i)\,\left(\frac{\partial \Psi}{\partial \phi}\right)^2&=&\frac{1}{4} r^2\sin^2\theta.
\end{eqnarray*}
The volume integral $\int d\mu(x_i) $ is given by $\int\sqrt{-g}\,d\tau dr d\theta d\phi$.  We assume that the right-hand side of the diagonal terms of Fisher information metric are representing the expectation values, which expressed by $\langle {\rm A} \rangle = \int d\mu(x_i)\,{\rm A}\Psi^2$.  Therefore, we have
\begin{eqnarray}
\int d\mu(x_i)\,\left[ -\left(1-\frac{r_s}{r}\right)\kappa_0'^2\Psi^2 - \left(\frac{\partial \Psi}{\partial \tau}\right)^2\right] &=& 0, \label{FIMtau'} \\
\int d\mu(x_i)\,\left[ \left(1-\dfrac{r_s}{r}\right)^{-1}\kappa_1'^2\Psi^2 - \left(\frac{\partial \Psi}{\partial r}\right)^2\right] &=& 0, \\
\int d\mu(x_i)\,\left[ \kappa_2'^2 r^2 \Psi^2 - \left(\frac{\partial \Psi}{\partial \theta}\right)^2\right] &=& 0, \\
\int d\mu(x_i)\,\left[ \kappa_3'^2 r^2\sin^2\theta \Psi^2 - \left(\frac{\partial \Psi}{\partial \phi}\right)^2\right] &=& 0 \label{FIMphi'}, 
\end{eqnarray}
where the factor $1/4$ is absorbed into the Lagrange multipliers $\kappa_i$.  We note that the unit vectors in the spherical polar coordinates are not constant, they depend on the position vector.  From the constraints of Fisher information metric, Eq.~(\ref{FIMtau'}-\ref{FIMphi'}), we can determine the gradients in the spatial dimensions are:
\begin{eqnarray*}
\nabla_\tau\Psi&=&-i\left(1-\frac{r_s}{r}\right)^{-1/2}\frac{\partial\Psi}{\partial \tau}\hat{\tau}\ , \\
\nabla_r\Psi&=&\left(1-\frac{r_s}{r}\right)^{1/2}\frac{\partial\Psi}{\partial r}\hat{r}\ ,\\
\nabla_\theta\Psi&=&\frac{1}{r}\frac{\partial\Psi}{\partial\theta}\hat{\theta}\ , \
\nabla_\phi\Psi=\frac{1}{r\sin\theta}\frac{\partial\Psi}{\partial\phi}\hat{\phi}\ ,
\end{eqnarray*}
Therefore we redefine the constraints of Fisher information metric, which are $\int d\mu(x_i)\,(\nabla_\tau\Psi)^2=\langle\kappa_0'^2\rangle$, $\int d\mu(x_i)\,(\nabla_r\Psi)^2=\langle\kappa_1'^2\rangle$, $\int d\mu(x_i)\,(\nabla_\theta\Psi)^2=\langle\kappa_2'^2\rangle$ and $\int d\mu(x_i)\,(\nabla_\phi\Psi)^2=\langle\kappa_3'^2\rangle$.  With a setted metric, we determine the rule of gradient and define the constraints which the probability distributions must satisfy.  In order to obtain the probability distributions by using the method called extreme physical information, we determine the objective function for the Fisher information metric for variations that vanish at the boundary, reads 

Then the objective function can be defined as
\begin{equation}
{\mathscr L}=\left(\nabla_\tau\Psi\right)^2+ \left(\nabla_r\Psi\right)^2+\left(\nabla_\theta\Psi\right)^2
 +\left(\nabla_\phi\Psi\right)^2-\alpha'^2\Psi^2\ ,\label{object01}
\end{equation}
where $\alpha'^2=\sum_{i=0}^3\kappa_i'^2$.  Now, extreming Eq.~(\ref{object01}) gives
\begin{widetext}
\begin{eqnarray}
\frac{1}{(1-r_s/r)}\frac{\partial^2\Psi}{\partial\tau^2}-\frac{1}{r^2} \frac{\partial}{\partial r}\left(r^2\left(1-\frac{r_s}{r}\right)
\frac{\partial\Psi}{\partial r}\right)
-\left[\frac{1}{r^2\sin\theta}\frac{\partial}{\partial\theta}\left(\sin\theta\frac{\partial\Psi}{\partial\theta}\right)
+\frac{1}{r^2\sin^2\theta}\left(\frac{\partial^2\Psi}{\partial\phi^2}\right)\right]
=\alpha'^2\Psi\ .\label{sHE01}
\end{eqnarray}
Considering the problem of finding solution of the form $\Psi(\tau, r,\theta,\phi) = T(\tau)R(r)\Theta(\theta)\Phi(\phi)$.  By separation variables, we obtain four differential equations
\begin{eqnarray}
\frac{d^2T(\tau)}{d\tau^2}+\eta'^2T(\tau)&=&0,\label{sEOMtau'}\\
\frac{1}{r^2}\frac{d}{d r}\left(r^2\left(1-\frac{r_s}{r}\right)
\frac{dR(r)}{d r}\right)+\left(\alpha'^2
+\frac{\eta'^2}{1-r_s/r}-\frac{\ell(\ell+1)}{r^2}\right)R(r)&=&0,\label{sEOMr'}\\
\frac{1}{\sin\theta}\frac{d}{d\theta}
\left(\sin\theta\frac{d\Theta(\theta)}{d\theta}\right)
+\left(\ell(\ell+1)-\frac{m^2}{\sin^2\theta}\right)\Theta(\theta)&=&0,
\label{sEOMtheta'}\\
\frac{d^2\Phi(\phi)}{d\phi^2}+m^2\Phi(\phi)&=&0.\label{sEOMphi'}
\end{eqnarray}
\end{widetext}
The temporal dependent solution to Eq.~(\ref{sEOMtau'}) is
\begin{equation}
T(\tau)= e^{-i\eta'\tau}\ , \ \eta'=0,\pm 1,\pm 2,\cdots. \label{sol tau}
\end{equation}
The azimuthal and polar dependents solutions are
\begin{eqnarray*}
\Phi(\phi)&=& e^{im\phi}\ , \ m=0,\pm 1,\pm 2,\cdots\ , \ -\ell\leq m\leq +\ell, \nonumber\\
\Theta(\theta)&=& \epsilon\sqrt{\dfrac{(2\ell+1)}{4\pi}\dfrac{(\ell-|m|)!}{((\ell+|m|)!)}} P_\ell^{m}(\cos\theta)\ , \ \ell=0,1,2,\cdots.\nonumber
\end{eqnarray*}
Finally, the differential equation Eq.~(\ref{sEOMr'}) describes the probability distributions in radial orientation, which does not have a simple solution.
When $1\gg r_s/r$, it can be approximated by the differential
equation that describes the probability distribution in spherical
Minkowski space, Eq.~(\ref{radialEQ}).

\subsection{Is there any connection to quantum physics?}
From the above derivations, we assign probability distributions to
describe space-times according to Fisher information metric.
Although we have the differential equation, Eq.~(\ref{sHE01}), to
describe the probability distribution in the Schwarzchild
space-time, we should like to understand the phenomena when we put a
test particle in a space-time. Consider the Fourier transform of
$\Psi(\tau,r,\theta,\phi)$ as follows:
\begin{eqnarray}
\Psi(\tau,r,\theta,\phi)&=&\dfrac{1}{(2\pi\hbar)^2}\int dE dp_r dp_\theta dp_\phi \Phi(E, p_r, p_\theta, p_\phi)\nonumber \\
& &\times  e^{-i(-E\tau + p_r r + p_\theta\theta + p_\phi\phi)/\hbar},\label{project02}
\end{eqnarray}
where the dimension of Planck's constant $h$ is the same as that of
the angular momentum, $\hbar = h/2\pi$. We substitute
Eq.~(\ref{project02}) into Eq.~(\ref{object01}) and get the
Lagrangian as
\begin{equation*}
\mathscr{L}=\left(-\dfrac{E^2}{\hbar^2(1-r_s/r)}+(1-\dfrac{r_s}{r})\dfrac{p_r^2}{\hbar^2}+\frac{L_\theta^2}{\hbar^2}+\frac{L_\phi^2}{\hbar^2}-\alpha'^2\right)\Psi^2,
\end{equation*}
where $E$ is the energy at infinity $r\to +\infty$, $L$ is the total
angular momentum, $p_\theta$ is the conjugate momentum to $\theta$,
$L_\theta$ is the angular momentum for the variable $\theta$, and
$p_\phi$ is the conjugate momentum to $\phi$, $L_\phi$ is the
angular momentum for the variable $\phi$.   A particle with non-zero
rest mass, $\mu\neq 0$, must satisfy a conservation equation, which
is~\cite{MTW}
\begin{equation*}
g_{\lambda\nu}p^\lambda p^\nu + \mu^2c^2=g^{\lambda\nu}p_\lambda p_\nu + \mu^2c^2=0,
\end{equation*}
where $c$ is the speed of light.  Therefore, we determine that
$\alpha^2$ should equal to $-\mu^2c^2/\hbar^2$. If there exists an
external potential field $V(r)$, then we let
$\alpha^2=-\mu^2c^2/\hbar^2+V(r)$.

An observer at rest on the equator of the Schwarzchild coordinate
system measures the total energy of the test particle when  it
passes him in its orbit, he gets $E_{\rm local}^2=(1-r_s/r)E^2$.
Therefore, we can define the local energy operator as $\hat{E}_{\rm
local}^2= -\hbar^2\nabla_\tau^2$.

\section{Conclusion}

In this article, we have proposed a method by identifying the metrics for physical space with the Fisher information metric, which determines the maximum number of distinguishable probabilities, to obtain the probability distributions for physical systems. From the relation between metric $g_{\mu\nu}$ and Fisher information matrix, we can postulate that there exist a certain configurations of probability distributions under certain constraints such that
\begin{equation*}
\int d\mu(x_i) \left(\frac{\partial\Psi}{\partial x^\mu}\right)^*\left(\frac{\partial\Psi}
{\partial x^\nu}\right)= \frac{1}{4} g_{\mu\nu},
\end{equation*}
where we need to consider the complex conjugate because wave
function could be complex, but $g_{\mu\nu}$ are real.

Next, we have generated the wave function from extreming the wave
function under certain constraints, such as normalization and localization of probabilities, via Lagrange multipliers.  From the derivation, we discover a differential equation for the wave function. For Minkowski metric, we obtained a differential equation in the Klein-Gordon equation form.  

After discussing the Minkowski metric, we considered the
Schwarzschild metric which describes the space-time due to a
spherical mass.
We obtained a differential equation for the wave function in the
Schwarzschild geometry. The differential equation,
Eq.~(\ref{sHE01}), can be solved by the method of separation of
variables. But the radial differential equation,
Eq.~(\ref{sEOMr'}), remaining to be solved.


\section*{Acknowledgements}
I would like to thank Professor Su-Long Nyeo for many valuable
discussions and the writting help.

\appendix
\section{Hydrogen Wave Function}
In this article, we can obtain the probability distributions in physical system according to identifying the the metrics of physical spaces with the Fisher metric.  Hence we suppose the solution of the hydrogenic Schr\"odinger equation should obey the constraints of Fisher information metric.
\begin{equation}
\int d\mu(x_i)\,\left(\dfrac{\partial\Psi}{\partial\xi^\mu} \right)^*\left(\dfrac{\partial\Psi}{\partial\xi^\nu}\right)=g_{\mu\nu}. \label{A1}
\end{equation}

First, we rewrite the Fisher information metric elements
\begin{eqnarray}
\int d\mu(x_i)\,\left(\frac{\partial \Psi}{\partial r} \right)^*\left( \frac{\partial \Psi}{\partial r} \right)&=&\int d\mu(x_i)\,\kappa_1^2|\Psi|^2,\label{rF}\\
\int d\mu(x_i)\,\left(\frac{\partial \Psi}{\partial \theta} \right)^*\left( \frac{\partial \Psi}{\partial \theta} \right)&=&\int d\mu(x_i)\,\kappa_2^2|\Psi|^2\, r^2,\label{tF}\\
\int d\mu(x_i)\,\left(\frac{\partial \Psi}{\partial \phi}\right)^*\left( \frac{\partial \Psi}{\partial \phi} \right)&=&\int d\mu(x_i)\,\kappa_3^2|\Psi|^2\, r^2\sin^2\theta,\label{pF}
\end{eqnarray}
where $\beta$, $\gamma$, and $\zeta$ are multipliers. The volume
integral $\int_X d\mu(x_i)$ is $\int\sqrt{-g}\,d\tau dr d\theta d\phi$.
The above equations are satisfied if and only if the probability
distribution $\rho$ is separable. The objective functions for the
Fisher information metric read
\begin{eqnarray*}
\mathscr{L}_r=(\nabla_r\Psi)^*(\nabla_r\Psi)-\kappa_1^2\Psi^2,\\
\mathscr{L}_\theta=(\nabla_\theta\Psi)^*(\nabla_\theta\Psi)-\kappa_2^2\Psi^2,\\
\mathscr{L}_\phi=(\nabla_\phi\Psi)^*(\nabla_\phi\Psi)-\kappa_3^2\Psi^2.
\end{eqnarray*}
The Euler-Lagrange equations read
\begin{eqnarray*}
\dfrac{1}{r^2}\left[\dfrac{\partial}{\partial r}\left(r^2\dfrac{\partial\Psi}{\partial r}\right)\right]+\kappa_1^2\Psi &=&0,\\
\dfrac{1}{r^2\sin\theta}\left[\dfrac{\partial}{\partial \theta}\left(\sin\theta\dfrac{\partial\Psi}{\partial \theta}\right)\right]+\kappa_2^2\Psi &=&0,\\
\dfrac{1}{r^2\sin^2\theta}\left[\dfrac{\partial^2\Psi}{\partial\phi^2}\right]+\kappa_3^2\Psi &=&0.
\end{eqnarray*}
We determine that $\kappa_1^2=(\alpha^2-\frac{\ell(\ell+1)}{r^2})$,
$\kappa_2^2=\frac{1}{r^2}[\ell(\ell+1)-\frac{m^2}{\sin^2\theta}]$, and
$\kappa_3^2=\frac{m^2}{r^2\sin^2\theta}$ by comparing with
Eqs.~(\ref{radialEQ}, \ref{polarEQ}, \ref{azimuthalEQ}).

Let us now check the relations of Fisher information metric,
Eqs.~(\ref{rF}, \ref{tF}, \ref{pF}) with the following wave function
of the hydrogen atom,
\begin{equation*}
\Psi_{322}=\frac{1}{162\sqrt{\pi}a^{3/2}}\frac{r^2}{a^2}e^{-r/3a}\sin^2\theta
e^{i2\phi}.
\end{equation*}
Then the elements of Fisher information metric are
\begin{eqnarray*}
\int drd\theta d\phi\,r^2\sin\theta\left(\frac{\partial\Psi_{322}}{\partial r}\right)^*
\left(\frac{\partial\Psi_{322}}{\partial r}\right)&=&\frac{1}{45 a^2}, \\
\int drd\theta d\phi\,r^2\sin\theta\left(\frac{\partial\Psi_{322}}{\partial\theta}\right)^*
\left(\frac{\partial\Psi_{322}}{\partial\theta}\right)&=&1, \\
\int drd\theta d\phi\,r^2\sin\theta\left(\frac{\partial\Psi_{322}}{\partial \phi}\right)^*\left(\frac{\partial\Psi_{322}}{\partial\phi}\right)&=&4.
\end{eqnarray*}
The $\theta$ and $\phi$ terms in the right-hand side of the elements
of Fisher information metric are
\begin{eqnarray*}
\int drd\theta d\phi\,r^2\sin\theta\Psi_{322}^*\dfrac{r^2}{r^2}\left[2(2+1)-\dfrac{2^2}{\sin^2\theta}\right]\Psi_{322}&=&1, \\
\int drd\theta d\phi\,r^2\sin\theta\Psi_{322}^*r^2\sin^2\theta\left[\dfrac{2^2}{r^2\sin^2\theta}\right]\Psi_{322}&=&4.
\end{eqnarray*}

Consequently, we show that the wave functions of hydrogen atom
satisfy the Fisher information metric, Eq.~(\ref{A1}).


\begin{thebibliography}{200}
\bibitem{Fisher} R. A. Fisher, Proc. Cambridge Philos. Soc. \textbf{22} (1925)~700.
\bibitem{Nagaoka} S. Amari and H. Nagaoka, {\it Methods of Information Geometry} (Oxford University Press, 2000).
\bibitem{Wootters} W. K. Wootters, Phys. Rev. D \textbf{23} (1981)~357.
\bibitem{Caves} S. L. Braunstein and C. M. Caves, Phys. Rev. Lett. \textbf{72} (1994)~3439.
\bibitem{Frieden2004} B. R. Frieden, {\it Science from Fisher Information} (Cambridge University Press, Cambridge, England, 2004).
\bibitem{Minkowski} H. Minkowski, "Raum und Zeit", Physikalische Zeitschrift {\bf 10} (1908/9)~75.
\bibitem{Calmet} X. Calmet and J. Calmet, Phys. Rev. E {\bf 71} (2005)~056109
\bibitem{Schrodinger}Erwin Schr\"odinger, Annalen der Physik {\bf 79}(4) (1926)~361.
\bibitem{Schwarzschild1916}K. Schwarzschild, Sitzungsber. Preuss. Akad. Wiss. Berlin (Math. Phys.) {\bf 1916} (1916)~189, ibid {\bf 1916} (1916)~424.
\bibitem{MTW} C. W. Misner, K. S. Thorne, J. A. Wheeler, {\it Gravitation} (W. H. Freeman and company, San Francisco, USA, 1973).
\end{thebibliography}
\end{document}